\documentclass[12pt,letterpaper]{article}

\usepackage{fixltx2e}
\usepackage{textcomp}
\usepackage{fullpage}
\usepackage{amsfonts}
\usepackage{verbatim}
\usepackage[english]{babel}
\usepackage{pifont}
\usepackage{color}
\usepackage{setspace}
\usepackage{lscape}
\usepackage{indentfirst}
\usepackage[normalem]{ulem}
\usepackage{booktabs}
\usepackage{natbib}
\usepackage{float}
\usepackage{latexsym}
\usepackage{url}
\usepackage{hyperref}
\usepackage{epsfig}
\usepackage{graphicx}
\usepackage{amssymb}
\usepackage{amsmath}
\usepackage{bm}
\usepackage{array}
\usepackage{mhchem}
\usepackage{ifthen}
\usepackage{caption}
\usepackage{hyperref}
\usepackage{amsthm}
\usepackage{amstext}

\usepackage{amsmath}
\usepackage{multirow}
\newtheorem{thm}{Theorem}

\newtheorem{obs}{Observation}

\DeclareMathOperator*{\argmin}{argmin}
\theoremstyle{definition}
\newtheorem{example}{Example}
\newtheorem{defn}{Definition}

\raggedright
\renewcommand{\section}[1]{%
\bigskip
\begin{center}
\begin{Large}
\normalfont\scshape #1
\medskip
\end{Large}
\end{center}}

\renewcommand{\subsection}[1]{%
\bigskip
\begin{center}
\begin{large}
\normalfont\itshape #1
\end{large}
\end{center}}

\renewcommand{\subsubsection}[1]{%
\vspace{2ex}
\noindent
\textit{#1.}---}

\renewcommand{\tableofcontents}{}

\bibpunct{(}{)}{;}{a}{}{,}  

\begin{document}
\begin{flushright}
Version dated: \today
\end{flushright}
\bigskip
\noindent Non-hereditary MDC trees

\bigskip
\medskip
\begin{center}

\noindent{\Large \bf Non-hereditary Minimum Deep Coalescence trees}
\bigskip

\noindent {\normalsize \sc Mareike Fischer$^1$, Martin Kreidl$^2$}\\
\noindent {\small \it 
$^1$Department of Mathematics and Computer Science, Ernst-Moritz-Arndt-University, Greifswald, 17489, Germany;\\
$^2$TWT GmbH Science \& Innovation, Munich, 80992, Germany}\\
\end{center}
\medskip
\noindent{\bf Corresponding author:} Mareike Fischer, Department of Mathematics and Computer Science, Ernst-Moritz-Arndt-University, Walther-Rathenau-Str. 47,
17489 Greifswald, Germany; E-mail: email@mareikefischer.de\\

\vspace{1in}

\subsubsection{Abstract} One of the goals of phylogenetic research is to find the species tree describing the evolutionary history of a set of species. But the trees derived from geneti data with the help of tree inference methods are gene trees that need not coincide with the species tree. This can for example happen when so-called deep coalescence events take place. It is also known that species trees can differ from their most likely gene trees. Therefore, as a means to find the species tree, it has been suggested to use subtrees of the gene trees, for example triples, and to puzzle them together in order to find the species tree. In this paper, we will show that this approach may lead to wrong trees regarding the minimum deep coalescence criterion (MDC). In particular, we present an example in which the optimal MDC tree is unique, but none of its triple subtrees fulfills the MDC criterion. In this sense, MDC is a non-hereditary tree reconstruction method.

\noindent (Keywords: minimum deep coalescence (MDC), parsimony, incomplete lineage sorting, subtree )\\

\vspace{1.5in}

In phylogenetics, there are basically two types of tree reconstruction problems: On the one hand, for given sequence data like DNA, RNA or protein data, one seeks to reconstruct the phylogenetic tree which explains these data best. There are several methods to do this, for example Maximum Parsimony (MP), Maximum Likelihood (ML), distance methods or Bayesian methods. It has long been known that different methods may lead to different trees (c.f. \cite{huson}, \cite{thatte}). However, even the same method can lead to different trees for the same set of species when it is applied to different genes (c.f. \cite{rosenberg}). In this case, we have a set of conflicting gene trees. It is therefore not trivial to find the species tree, i.e. the tree describing the true evolutionary history of the species under investigation. So the second tree reconstruction problem is concerned with estimating the species tree from a set of input gene trees.

Different gene trees based on the same species tree can have various known causes. For example, horizontal gene transfer may lead to gene trees which contradict each other. In this case, a phylogenetic network (\cite{kunin}) would be more appropriate than a tree in order to describe the underlying process. But there are also cases when evolution is treelike and still some gene trees do not coincide with the species tree. This can happen when so-called incomplete lineage sorting takes place. For example, consider Figure \ref{mapping}: If tree $S$ on the right hand side of this figure is the underlying species tree and $T$ is a gene tree, then there are two instances (highlighted by the small arrows) where the coalescence of individual lineages of $T$ happens further towards the root of the tree than suggested by $S$. This means that we have incomplete lineage sorting  whenever two ancestral lineages in the gene tree fail to coalesce before (looking backwards in the tree) more recent speciation events take place (\cite{maddison}). This problem, which is more likely when branches of the species tree are short and the population size of the respective species is large, is therefore also known as deep coalescence. One way to estimate the species tree is thus to seek the Minimum Deep Coalescence (MDC) tree, i.e. the (not necessarily unique) tree which requires a minimum number of deep coalescence events to explain the given gene trees.

The concept of MDC was first introduced by \cite{maddison}, who also realized that this concept is in some sense a parsimony concept. Note that parsimony in phylogenetics usually refers to Maximum Parsimony (MP), a method of finding the tree which requires the smallest number of mutations to explain a given sequence alignment by, simply speaking, finding a consensus between all input sites in the alignment (\cite{bryant}). What Maddison means, though, is similar in the sense that a consensus is aimed at, but in this case a consensus between different gene trees.  However, finding a consensus usually comes with a price, and MP therefore has various drawbacks (c.f. \cite{felsenstein}, \cite{fischer}, \cite{thatte}). One of those drawbacks is sometimes referred to as non-heredity: In fact, for a given DNA alignment, it can happen that you have a unique MP tree, but non of its subtrees is MP for the corresponding subalignment (c.f. \cite{fischer}). So in this regard, it is an  interesting question to see if MDC as a parsimony method in a different context, is hereditary or non-hereditary.

In particular, it has been proposed (c.f. \cite{rosenberg}) that in order to estimate the species tree, instead of estimating gene trees for each set of genes and estimating the species tree from those, smaller gene trees like e.g. triples, i.e. trees on only three species at a time, should be reconstructed and then puzzled together in order to build the estimated species tree. This method is well-known in phylogenetics and often referred to as tree puzzling (\cite{haeseler}). 

Note that MDC is NP-hard (\cite{zhang}), i.e. it is hard to find a tree which minimizes the sum of the MDC scores of the input gene trees. However, this does not immediately imply that MDC is non-hereditary: even if it were hereditary, i.e. even if each MDC tree always had an MDC subtree, it would not be clear which one to choose. 

Therefore, in this paper we investigate the question whether MDC trees are hereditary. In particular, we will analyze whether a unique MDC tree necessarily has at least one MDC subtree. We answer this question negatively even for four taxa: We present an example of a set of gene trees on four taxa which have a unique MDC tree, which in turn has no subtree of size three that is MDC for the input gene trees' subtrees of size three. This proves that MDC trees cannot be reconstructed from subtrees of smaller sizes. Moreover, the main idea used in this paper to derive our result is similar to the ideas presented in \cite{fischer} in the sense that we construct a system of inequalities based on a formula to calculate the MDC score proven by \cite{nakleh} and check if the solution space is empty or not. We believe that this approach can also be useful for tackling other questions concerning MDC.

\section{Preliminaries}
We need to introduce some concepts and notations before we can present our results. In this paper, we discuss so-called {\em rooted binary phylogenetic trees} on a set $X$ of species or taxa. Recall that a tree is a connected acyclic graph, and a phylogenetic tree has its leaves labelled by the names of the underlying set of species. Such a tree is rooted and binary if all internal nodes have degree 3, except for one specific node representing the most recent common ancestor of all species under investigation, which is called the {\em root} and has degree 2. When there is no ambiguity, we will refer to rooted binary phylogenetic trees only as trees for short. For a tree $T$ and a node $v$ in $T$, we denote by $T(v)$ the {\em clade} of $T$ rooted at $v$. The leaves of this clade are called {\em cluster}, and this cluster is denoted by $C_T(v)$. When the node $v$ on which a cluster $t$ is pending is not explicitly stated, such a leaf set can also be denoted by $\mathcal{L}(t)$. In a tree $T$ on species set $X$, the {\em most recent common ancestor}, or {\em MRCA} for short, of a set $y \subseteq X$ of species is a node $u$ in $T$ such that all leaves in $Y$ are descendants of $u$ and all other nodes with this property are closer to the root than $u$. So the MRCA of a cluster $C_T(v)$ is $v$. In the special case where $Y=X$, the MRCA is the root of $T$.

When we consider a subtree $T'$ of a tree $T$, this means we are restricting the leaf set $X$ to the leaf set $Y \subset X$ of $T'$, i.e. we delete all leaves which are not in $X$ as well as edges leading to these leaves and, if this implies that the root no longer has degree 2, we also delete the root and the remaining edge leading to the root. Then we suppress all resulting nodes of degree 2 (other than possibly the MRCA of $Y$ in case that the original root has been deleted, because this MRCA is then the new root of $T'$) in order to obtain a binary tree again. In case a tree $T'$ is derived from $T$ in this manner, we write $T' = T|_{Y}$, i.e. $T'$ is regarded as the restriction of $T$ to $Y$. Note that a clade is a particular kind of subtree, namely a subtree which is induced by a node $v$ of $T$. For example, consider tree $T_1$ from Figure \ref{alltrees}. Then, considering Figure \ref{allsubtrees} $T_{\{1,2,3\}}=\tilde{T_1}$,  $T_{\{1,2,4\}}=\hat{T_1}$,  $T_{\{1,3,4\}}=\bar{T_1}$ and  $T_{\{2,3,4\}}=\dot{T_1}$, but only $\tilde{T_1}$ is a clade of $T_1$, and thus $T_1$ contains the cluster $\{1,2,3\}$, but not, say, $\{1,2,4\}$. 

Note that when a collection of trees is given, this collection may actually be a multiset, because some trees could occur more than once (e.g. if various genes lead to the same gene tree). We refer to such a multiset of $m$ (gene) trees as $G_m$, and we denote with $G$ the corresponding (simple) set which contains all trees of $G_m$ exactly once. When such a set of trees on the same set of species is given, their minimum deep coalescence tree, or MDC tree for short, can be defined as the tree minimizing the number of so-called {\em extra lineages} (\cite{maddison}). In order to define these extra lineages, we follow the approach by \cite{nakleh} and start by introducing the following mapping. For a given gene tree $T$ and a given species tree $S$ on the same taxon set $X$, we fit $T$ into $S$ as follows: 
\begin{enumerate}\item Each taxon of $T$ is mapped into the corresponding taxon in $S$. \item If $v'$ is the most recent common ancestor of a cluster $C_T(v)$ of $T$ in $S$, and if $u'$ is the parent node of $v'$ in $S$, then $v$ is mapped to some point on the edge $(u',v')$ except $u'$. \item If a node $u$ is an ancestor of a node $v$ in $T$, then for their images in $S$, say $p_u$ and $p_v$, we have that $p_u$ is an ancestor of $p_v$ in $S$, too.
\end{enumerate}

\begin{figure}
\center
\scalebox{1.15}{ \includegraphics{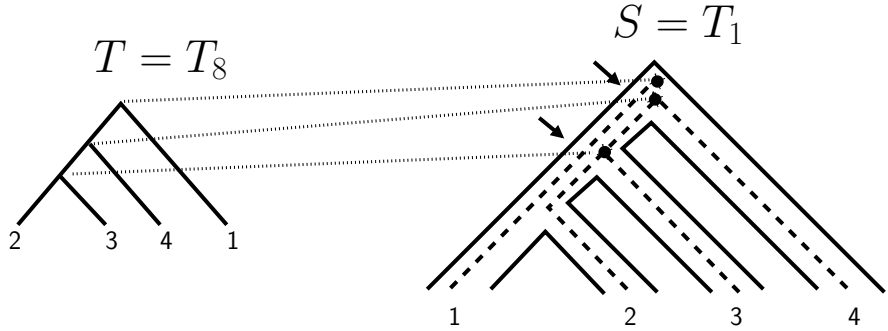}}
\caption{The mapping of the gene tree $T_8$ from Figure \ref{alltrees} into the species tree $T_1$. All leaves of the gene tree are mapped into the corresponding leaves of the species tree; for all inner nodes the mapping is indicated by the dotted lines. In total, two extra lineages are needed and highlighted by the small arrows. }\label{mapping}
\end{figure}

This mapping is illustrated by Figure \ref{mapping}. Here, a gene tree $T$ is mapped into a species tree $S$. Only the mapping of the inner nodes of $T$ is depicted by the dotted lines -- all taxa are mapped into their respective counterparts of $S$.

Now, given a gene tree $T$ and a species tree $S$ and the above mapping, \cite{nakleh} define the number of extra lineages of a branch in $S$ as the number of lineages of $T$ that exit the branch (when looking backwards in time, i.e. towards the root) -1. The number of extra lineages of $T$ on $S$ is then simply the sum of extra lineages over all edges. In Figure \ref{mapping}, there are two extra lineages, which are highlighted by the small arrows. Here, looking backwards in time and leaving an edge of the species tree, at both arrows there are two dashed lineages of $T$ present. As one lineage would be the ideal case, both edges with two lineages give an extra lineage, which means there are in total two extra lineages in this example.

Now these extra lineages play a fundamental role as MDC trees can be defined with their help: For a set $G$ of gene trees, the tree $S$ which requires the minimum number of extra lineages in total, i.e. when the sum over all trees in $G$ is taken, is called {\em MDC tree}. Luckily, while it is hard to find an MDC tree, it is easy to calculate the number of extra lineages of a tree $S$ with the help of Theorem 2 of \cite{nakleh}:

\begin{thm}[Theorem 2 of \cite{nakleh}] Let $T$ be a gene tree and $S$ a species tree. Let $(u',v')$ be an edge in $S$. Denote by $t_1,t_2,\ldots,t_k$ all maximal clades of $T$ such that $\mathcal{L}(t_i) \subseteq C_S(v')$ for $i \in \{1,\ldots,k\}$. (Here, $\mathcal{L}(t_i)$ denotes the leaf set of $t_i$). Then, the number of extra lineages in $(u',v')$ is $k-1$.
\end{thm}

This theorem gives a simple formula for calculating the number of extra lineages of each edge of $S$, and thus -- by taking the sum over all edges -- of $S$. We denote by $l(T,S)$ the minimum number of extra lineages a tree $T$ needs in a species tree $S$ and call $l(T,S)$ the {\em MDC score} of $T$ in $S$. 

Simply put, for each branch we just have to count how many maximal clades of $T$ have their leaf set contained in the cluster of $S$ pending on the edge under investigation, and then take $-1$. The -1 is due to the fact that one lineage of the gene tree is needed per lineage of the species tree, but all additional lineages are extra lineages. In Figure \ref{mapping}, we have already seen that the total number of extra lineages is 2. However, we could have derived this with the above theorem es follows: $S$ has clusters $(1,2)$, $(1,2,3)$ as well as the trivial clusters $(1)$, $(2)$, $(3)$, $(4)$ and $(1,2,3,4)$, which all trees on taxon set $X=\{1,2,3,4\}$ have. $T$, on the other hand, has clusters: $(2,3)$, $(2,3,4)$ and the trivial ones. Now for each cluster of $S$ we check how many maximal clusters of $T$ are contained in this cluster. Note that for the trivial clusters, the answer of course is always 1, because they are contained in both trees. So we focus on the other clusters and start with $(1,2)$. We find that $(1)$ and $(2)$ of $T$ are contained in this cluster, so this gives $k=2$ for the edge inducing cluster $(1,2)$. Similarly, $(1,2,3)$ contains clusters $(1)$ and $(2,3)$ and thus again, we have $k=2$. So in total we have $(2-1) + (2-1)=2$ extra lineages, which confirms our earlier result.

Table \ref{tab5} summarizes the number of extra lineages for all possible gene tree / species tree combinations on all trees on four taxa as depicted by Figure \ref{alltrees}. Note that this matrix is {\em not} symmetric. For example, we have $l(T_1,T_8)=3$ and $l(T_8,T_1)=2$ (both highlighted in bold), so the roles of the gene tree and the species tree are {\em not} interchangeable.

\begin{table}\begin{tabular}{ |cc|ccccccccccccccc|  }   \hline   & &\multicolumn{15}{|c|}{species tree $S$ }Ê\\   \parbox[t]{2mm}{\multirow{15}{*}{\rotatebox[origin=c]{90}{gene tree $T$}}}
& &$T_1$ &  $T_2$ &  $T_3$ &  $T_4$ &  $T_5$ &  $T_6$ &  $T_7$ &  $T_8$ &  $T_9$ &  $T_{10}$ &  $T_{11}$ &  $T_{12}$ &  $T_{13}$ &  $T_{14}$ &  $T_{15}$ \\ \hline  &$T_1$ & 0 & 1 & 1 & 3 & 2 & 3 & 1 & {\bf 3} & 2& 3 & 3 & 3 & 1 & 2 & 2 \\  &$T_2$ & 1 & 0 & 2 & 3 & 1 & 3 & 2 & 3 & 1& 3 & 3 & 3 & 1 & 2 & 2 \\  &
$T_3$ & 1 & 3 & 0 & 1 & 3 & 2 & 1 & 3 &3& 3 &2 & 3 & 2 & 1 & 2 \\ &
$T_4$ & 2 & 3 & 1 & 0 & 3 & 1 & 2 & 3 & 3& 3 & 1 & 3 & 2 & 1 & 2 \\  &
$T_5$ & 3 & 1 & 3 & 2 & 0 & 1 & 3 & 3 & 1& 3 & 2 & 3 & 2 & 2 & 1 \\  &
$T_6$ & 3 & 2 & 3 & 1 & 1 & 0 & 3 & 3 & 2& 3 & 1 & 3 & 2 & 2 &1 \\  &
$T_7$ & 1 & 3 & 1 & 3 & 3 & 3 & 0 & 1 & 3& 2 & 3 & 2 &2 & 2 & 1 \\  &
$T_8$ & {\bf 2} & 3 & 2 & 3 &3 & 3 & 1 & 0 & 3& 1 & 3 & 1 & 2 & 2 & 1 \\  &
$T_9$ & 3 & 1 & 3 & 3 & 1 & 3 &3 & 2 & 0& 1 & 3 & 2 & 2 & 1 & 2 \\  &
$T_{10}$ & 3 & 2 & 3 & 3 & 2 & 3 & 3 & 1 & 1& 0 & 3 & 1 & 2 & 1 & 2 \\  &
$T_{11}$ & 3 & 3 & 3 & 1 &3 & 1 & 3 & 2 & 3&2 & 0 & 1 & 1 & 2 & 2 \\  &
$T_{12}$ & 3 & 3 & 3& 2 &3 & 2 & 3 & 1 & 3& 1 & 1 & 0 & 1 & 2 & 2 \\  &
$T_{13}$ & 1 & 1 & 2 & 2 & 2 & 2 & 2 & 2 & 2& 2 & 1 & 1 & 0 & 2 & 2 \\  &
$T_{14}$ & 2 & 2 & 1 & 1 & 2 & 2 & 2 & 2 & 1& 1 & 2 & 2 & 2 & 0 & 2 \\  &
$T_{15}$ & 2 & 2 & 2 & 2 & 1 & 1 & 1 & 1 & 2&2 & 2 & 2 & 2 & 2 & 0 \\  
\hline  \end{tabular} \caption{Ranging over all trees from Figure \ref{alltrees}, this table gives the corresponding MDC scores $l(T,S)$, where $T$ denotes the assumed gene tree and $S$ the species tree. The values $l(T_1,T_8)=3$ and $l(T_8,T_1)=2$ are highlighted in bold to demonstrate that this matrix is not symmetric. }
\label{tab5}
\end{table}

We now derive the following definition of MDC trees.

\begin{defn} \label{char} A tree $S$ is an MDC tree for a multiset $G_m=\{T_1,\ldots,T_m\}$ of gene trees on a taxon set $X$ if and only if $$S=\argmin\limits_{S' \in \mathcal{T}} \sum\limits_{e \in E(S') } \sum\limits_{i=1}^m k_{T_i}(e) = \argmin\limits_{S' \in \mathcal{T}} \sum\limits_{i=1}^m l(T_i,S')=\argmin\limits_{S' \in \mathcal{T}} \sum\limits_{T \in G} x_T\cdot l(T,S').$$ Here, $\mathcal{T}$ denotes all rooted binary phylogenetic trees on taxon set $X$, $E(S')$ denotes the edge set of tree $S'$, $k_{T_i}(e)$ denotes the number of extra lineages edge $e$ of $S$ needs when $T_i$ is mapped to $S'$, $G$ denotes the simple set induced by $G_m$ and $x_T$ denotes the number of times tree $T$ is contained in $G_m$. 
\end{defn}

We are now in a position to state our results.

\section{Results}  
We will now introduce heredity of a species tree, which is the main concept of this paper. 

\begin{defn}
Let $S$ be an MDC tree for a multiset $G_m=T_1,\ldots,T_m$ of gene trees. Now we consider the set $G_m^i=\{T|_Y: T \in G\mbox{ and }Y= X \setminus \{i\}\}$, i.e. the set containing all subtrees of the trees in $G_m$ which result from deleting leaf $i$. Then, $S$ is called hereditary if there is an $i \in X$ such that $S|_{X \setminus \{i\}}$ is an MDC tree for $G_m^i$.
\end{defn}

Informally speaking, an MDC tree $S$ is hereditary if it contains at least one MDC subtree. This way, MDC trees could be traced back to smaller MDC trees by deleting one leaf at a time (however, it would not be clear which one to delete). We now state our main observation before we prove it subsequently.

\begin{obs} There exist gene trees $T_1,\ldots,T_m$ on a taxon set $X$ such that their MDC tree $S$ is unique, but such that $S|_Y$ is not an MDC tree for $T_1|_Y,\ldots,T_m|_Y$ for all subsets $Y$ with $Y\subset X$, $|Y|=|X|-1$. This implies that removing one taxon from the analysis changes the topology of the MDC tree.
\end{obs}

Next we describe our approach of constructing an example to prove the observation. We believe that this approach can be useful for proving or disproving other statements regarding MDC.

\begin{figure}
\center
\scalebox{1}{ \includegraphics{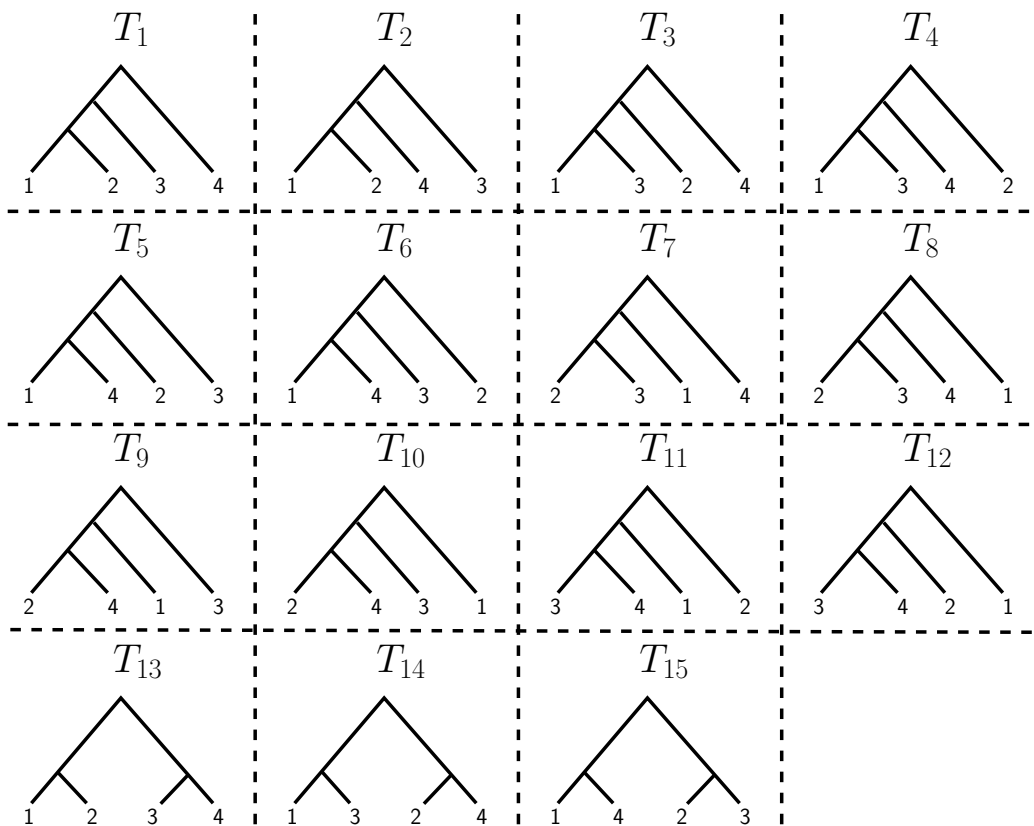}}
\caption{All 15 rooted binary trees on taxon set $X=\{1,2,3,4\}$. The tree shape of trees $T_{13}$,  $T_{14}$ and  $T_{15}$ is called balanced. Note that there are only three balanced trees, but twelve unbalanced ones. }\label{alltrees}
\end{figure}

\begin{figure}
\center
\scalebox{1}{ \includegraphics{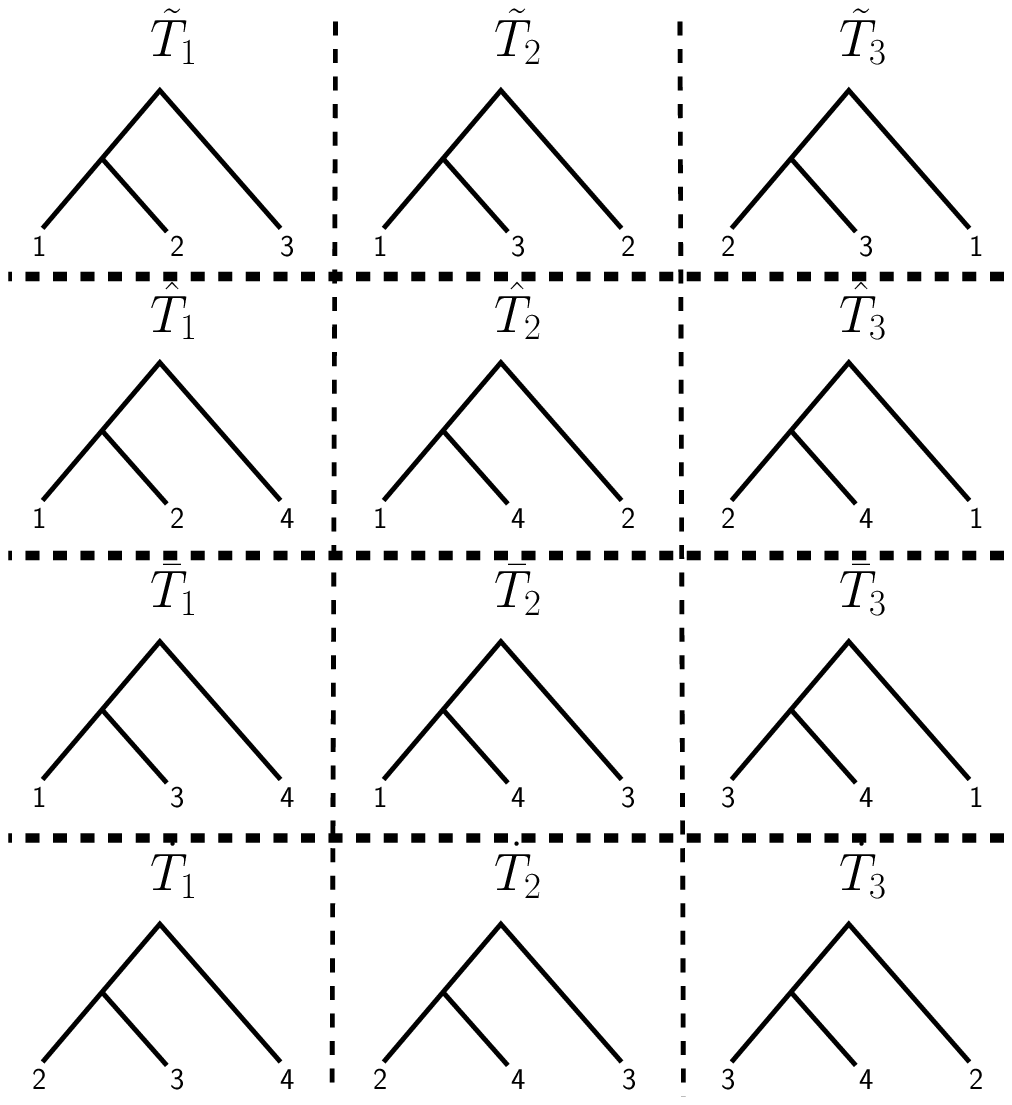}}
\caption{We consider all trees from Figure \ref{alltrees} on taxon set $X=\{1,2,3,4\}$. Now we restrict the taxon set on $\tilde{Y}=\{1,2,3\}$, i.e. $Y$ is derived from $X$ by deleting taxon 4, and consider all possible rooted trees on $Y$. This leads to $\tilde{T_1}$, $\tilde{T_2}$ and $\tilde{T_3}$. We repeat this for the subsets $\hat{Y}$ (deletion of leaf 3), $\bar{Y}$ (deletion of leaf 2) and $\dot{Y}$ (deletion of leaf 1) and receive the possible trees depicted in lines 2 -- 4 of the above table. } \label{allsubtrees}
\end{figure}

Note that by Definition \ref{char} a tree $S$ is an MDC tree for $G_m=T_1,\ldots,T_m$, where all $T_i$ are trees on a common taxon set $X$, if it minimizes $ \sum\limits_{T \in G} l(T,S')$ over all trees $S'$ on $X$, so $S$ is an MDC tree if an only if  \begin{equation}\label{ineq1} \sum\limits_{i =1}^{m} l(T_i,S) \leq  \sum\limits_{i =1}^{m} l(T_i,S') \mbox{  \hspace{0.5cm} for all trees $S'\neq S$ on taxon set $X$.}\end{equation} So for example, if we look at $X=\{1,2,3,4\}$, there are 15 rooted binary phylogenetic trees $T_1,\ldots, T_{15}$, which are all depicted by Figure \ref{alltrees}. If we want one of them, say $T_1$, to be an MDC tree, (\ref{ineq1}) gives us 14 inequalities which a multiset of gene trees $G_m$ would have to fulfill, because $T_1$ has to be at least as good as any of the other 14 trees. If we want $T_1$ to be the unique MDC tree of $G_m$, all inequalities in (\ref{ineq1}) are strict.

By the same reasoning, however, if we do not want a tree to be an MDC tree, only one of the inequalities needs to fail (i.e. at least one other tree needs to give a strictly lower MDC score than the tree under investigation, but not necessarily all of them). Consequently, if we are searching now for a tree $S$ on $X$ which has no MDC subtree on any $Y\subset X$ with $|Y|=|X|-1$, we require \begin{equation}\label{ineq2}  \sum\limits_{i=1}^m l(T_i|_Y,S|_Y) > \min\limits_{S' \in \mathcal{T} }\left\{ \sum\limits_{i=1}^m l(T_i|_Y,S'|_Y)\right\}  \mbox{ for all $Y \subset X$: $|Y|=|X|-1$,}\end{equation}

where $\mathcal{T}$ denotes the set of all binary phylogenetic trees on taxon set $X$. So for a given set $X$ of taxa, it remains to check if there is a tree $S$ which fulfills both (\ref{ineq1}) and (\ref{ineq2}) simultaneously. We answer this affirmatively for $X=\{1,2,3,4\}$ with the following example.

\begin{example} We consider the case $X=\{1,2,3,4\}$, for which 15 trees exist as depicted by Figure \ref{alltrees}. We consider the multiset $G_m$ which consists of 11 copies of $T_2$, 10 copies of $T_{4}$, 2 copies of $T_6$, 3 copies of $T_7$ and 3 copies of $T_{15}$, i.e. we have $x_2=11$, $x_4=10$, $x_6=2$, $x_7=3$, $x_{15} =3$ and $x_i=0$ for all other $i$. Note that $G_m$ contains no copy of $T_1$, but still $T_1$ can be shown to be the unique MDC tree of $G_m$. This can be seen by looking at Table \ref{tab1}, whose 3\textsuperscript{rd} column contains the MDC scores of all trees -- $T_1$ gives the unique minimum value. This shows why MDC is truly just a consensus method: It does not give back the majority tree, but instead it gives a compromise of the input trees, which may well not be contained in the data. However, in this example, all subtrees of $T_1$ are not MDC trees: When considering $G_m|_Y$ with $Y\subset X$ and $|Y|=3$, the corresponding subtree of $T_1$ is not MDC for $Y$, as can also be seen in Table \ref{tab1}. Note that our construction of $G_m$ with $|G_m|=29$ as summarized in Table \ref{tab1} is minimal in the sense that there is no smaller set $G_m$ on 4 taxa with the same properties, i.e. such that $T_1$ is uniquely MDC but non of its subtrees is MDC and that additionally $T_1$ is not contained in $G_m$. We verified this by an exhaustive search through all hypothetical smaller solutions with the help of a computer algebra system (calculations not shown).
\end{example}

\begin{example} In order to show that our result in Example 1 does not depend on the tree shape, we constructed the following example: $G_m$ now consists of 2 copies of $T_1$, 2 copies of $T_{12}$, 1 copy of $T_{14}$ and 1 copy of $T_{15}$. Here, $T_{13}$, whose tree shape is balanced (as opposed to that of $T_1$ as considered in the first example) is the unique MDC tree which again has no MDC subtrees. Table \ref{tab2} summarizes these findings. Moreover, note that $T_{13}$ is not contained in $G_m$. As with Example 1, this construction with $|G_m|=6$ is minimal: There is no smaller set $G_m$ such that $T_{13}$ is the unique MDC tree, but has no MDC subtrees and is not contained in $G_m$. However, when comparing Example 2 with Example 1, we conclude that while such constructions are possible for both tree shapes on 4 taxa, those for the more balanced tree shape require fewer trees in $G_m$ than for the other tree shape. 
\end{example}

\begin{example} While we find Example 1 and 2 interesting exactly because the MDC tree under consideration does not occur in $G_m$ at all, this might lead to the wrong conclusion that the non-heredity is caused by this fact. Therefore, we repeated our calculations and searched for a case where the opposite is true, namely that $T_1$ is the unique MDC tree and has no MDC subtrees but is also the most frequent tree in $G_m$. Table \ref{tab3} summarizes a minimal example for this scenario. Here, $G_m$ consists of 8 copies of $T_1$, 6 copies of $T_{2}$, 1 copy of $T_{3}$, 7 copies of $T_4$, 7 copies of $T_6$ and 4 copies of $T_{15}$. Here, $|G_m|=33$, so this example requires slightly more trees in $G_m$ than Example 1, where $T_1$ was not contained in $G_m$. This might be due to the fact that if the support for $T_1$ is stronger than that for any other tree in $G_m$, then the signal induced by $T_1$ for the subtrees is also strong, which is why more trees are needed to annihilate this signal of the subtrees. However, this example is also interesting for another reason: Here, not only are the subtrees of $T_1$ each outperformed by some subtrees of the other $T_i$, but there is in fact one tree, namely $T_6$, of which {\em all} subtrees outperform the corresponding subtrees of $T_1$ concerning MDC, but still $T_6$ is outperformed by $T_1$. So the signal given by the trees of size 3 leads to an entirely different conclusion than the signal induced by the trees of size 4. 
\end{example}

\begin{example} Our last example repeats the construction of Example 3 for the case that the unique MDC tree is $T_{13}$, and thus the tree shape is the balanced one. Here, $G_m$ consists of 3 copies of $T_1$, 1 copy of $T_{10}$, 2 copies of $T_{12}$, 1 copy of $T_{14}$ and 1 copy of $T_{15}$. The total number of trees in $G_m$ of 8 is minimal in the sense that for a smaller multiset of trees, it is not possible that $T_{13}$ is the unique MDC tree with no MDC subtree and such that $T_{13}$ occurs most frequently in $G_m$. Again, as in the comparison of Example 1 and Example 2, we realize that we need significantly fewer trees in order to construct such an example for the balanced tree shape than for the other one. Moreover, considering Example 2, we conclude that we need slightly more trees in $G_m$ if we make the signal for $T_{13}$ strong than if $T_{13}$ is not present at all. This coincides with the findings explained in Example 3 for the other tree shape.
\end{example}

\begin{table}\begin{tabular}{ |c||c|c|c|c|c|c|  }  \hline $T_i$& $x_i$ & {\scriptsize $l(G_m, T_i)$} & {\scriptsize $l(G_m|_{\{1,2,3\}},T_i|_{\{1,2,3\}})$}&{\scriptsize $l(G_m|_{\{1,2,4\}},T_i|_{\{1,2,4\}})$}&{\scriptsize$l(G_m|_{\{1,3,4\}},T_i|_{\{1,3,4\}})$}&{\scriptsize$l(G_m|_{\{2,3,4\}},T_i|_{\{2,3,4\}})$} \\\hline\hline $T_1$ & 0 & {\bf 46} &18&15&16&23 \\ \hline $T_2$ & 11 & 49 &18&15&{\bf 13}&18 \\\hline $T_3$ & 0 & 47 &{\bf 17}&15&16&23 \\ \hline $T_4$ & 10 & 50 &{\bf 17}&{\bf 14}&16&{\bf 17} \\ \hline $T_5$ &0 & 55 &18&{\bf 14}&{\bf 13}&18 \\ \hline $T_6$ & 2 & 55 &{\bf 17}&{\bf 14}&{\bf 13}&{\bf17} \\ \hline $T_7$ & 3 & 51 &23&15&16&23 \\ \hline $T_8$ & 0 & 75 &23&29&29&23 \\ \hline $T_9$ & 0 & 60 &18&29&{\bf 13}&18 \\ \hline $T_{10}$ & 0 & 81 &23&29&29&18 \\ \hline $T_{11}$ & 0 & 60 &{\bf 17}&{\bf 14}&29&{\bf 17} \\ \hline $T_{12}$ & 0 & 81 &23&29&29&{\bf 17} \\ \hline $T_{13}$ & 0 & 47 &18&15&29&{\bf 17 }\\ \hline $T_{14}$ & 0 & 48 &{\bf 17}&29&16&18 \\ \hline $T_{15}$ & 3 & 47 &23&{\bf 14}&{\bf 13}&23 \\       \hline \end{tabular} \caption{Example 1: Here, we consider a multiset of trees $G_m$, which consists of 11 copies of $T_2$, 10 copies of $T_{4}$, 2 copies of $T_6$, 3 copies of $T_7$ and 3 copies of $T_{15}$. The total number of 29 is minimal, i.e. $G_m$ is a minimal multiset, such that $T_1$ is the unique MDC tree but has no MDC subtrees of size 3 and $T_1 \not\in G_m$. }
\label{tab1}
\end{table}

\begin{table}\begin{tabular}{ |c||c|c|c|c|c|c|  }  \hline $T_i$& $x_i$ & {\scriptsize $l(G_m, T_i)$} & {\scriptsize $l(G_m|_{\{1,2,3\}},T_i|_{\{1,2,3\}})$}&{\scriptsize $l(G_m|_{\{1,2,4\}},T_i|_{\{1,2,4\}})$}&{\scriptsize$l(G_m|_{\{1,3,4\}},T_i|_{\{1,3,4\}})$}&{\scriptsize$l(G_m|_{\{2,3,4\}},T_i|_{\{2,3,4\}})$} \\\hline\hline $T_1$ & 2 & 10 &4&4&{\bf 3}&{\bf 3} \\ \hline $T_2$ & 0 & 12 &4&4&5&5 \\\hline $T_3$ & 0 & 11 &5&4&{\bf 3}&{\bf 3} \\ \hline $T_4$ & 0 &13 &5&5&{\bf 3}&4 \\ \hline $T_5$ &0 &13 &4&5&5&5 \\ \hline $T_6$ & 0 & 13 &5&5&5&4 \\ \hline $T_7$ & 0 & 11 &{\bf 3}&4&{\bf 3}&{\bf 3} \\ \hline $T_8$ & 0 & 11 &{\bf 3}&{\bf 3}&4&{\bf 3} \\ \hline $T_9$ & 0 & 13 &4&{\bf 3}&5&5 \\ \hline $T_{10}$ & 0 & 11 &{\bf 3}&{\bf 3}&4&5 \\ \hline $T_{11}$ & 0 & 12 &5&5&4&4\\ \hline $T_{12}$ & 2 & 10 &{\bf 3}&{\bf 3}&4&4 \\ \hline $T_{13}$ & 0 & {\bf 8} &4&4&4&4\\ \hline $T_{14}$ & 1 & 10 &5&{\bf 3}&{\bf 3}&5 \\ \hline  $T_{15}$ & 1 & 10 &{\bf 3}&5&5&{\bf 3} \\       \hline \end{tabular}\caption{Example 2: Here, we consider a multiset of trees $G_m$, which consists of 2 copies of $T_1$, 2 copies of $T_{12}$, 1 copy of $T_{14}$ and 1 copy of $T_{15}$. The total number of 6 is minimal, i.e. $G_m$ is a minimal multiset, such that $T_{13}$ is the unique MDC tree but has no MDC subtrees of size 3 and $T_{13} \not\in G_m$. }
\label{tab2}
\end{table}

\begin{table}
\begin{tabular}{ |c||c|c|c|c|c|c|  }  \hline $T_i$& $x_i$ & {\scriptsize $l(G_m, T_i)$} & {\scriptsize $l(G_m|_{\{1,2,3\}},T_i|_{\{1,2,3\}})$}&{\scriptsize $l(G_m|_{\{1,2,4\}},T_i|_{\{1,2,4\}})$}&{\scriptsize$l(G_m|_{\{1,3,4\}},T_i|_{\{1,3,4\}})$}&{\scriptsize$l(G_m|_{\{2,3,4\}},T_i|_{\{2,3,4\}})$} \\\hline\hline $T_1$ & 8 & {\bf 50} &19&18&17&20 \\ \hline $T_2$ & 6 & 54 &19&18&{\bf 16}&27 \\\hline $T_3$ & 1 & 56 &{\bf 18}&18&17&20 \\ \hline $T_4$ & 7 &58 &{\bf 18}&{\bf 15}&17&{\bf 19 }\\ \hline $T_5$ &0 &57 &19&{\bf 15}&{\bf 16}&27 \\ \hline $T_6$ & 7 & 55 &{\bf 18}&{\bf 15}&{\bf 16}&{\bf 19 }\\ \hline $T_7$ & 0 & 60 &29&18&17&20 \\ \hline $T_8$ & 0 & 91 &29&33&33&20 \\ \hline $T_9$ & 0 & 68 &19&33&{\bf 16}&27 \\ \hline $T_{10}$ & 0 & 95 &29&33&33&27 \\ \hline $T_{11}$ & 0 & 66&{\bf 18}&{\bf 15}&33&{\bf 19}\\ \hline $T_{12}$ & 0 & 95 &29&33&33&{\bf 19} \\ \hline $T_{13}$ & 0 & 52 &19&18&33&{\bf 19}\\ \hline $T_{14}$ & 0 & 58 &{\bf 18}&33&17&27 \\ \hline  $T_{15}$ & 4 & 51 &29&{\bf 15}&{\bf 16}&20 \\       \hline \end{tabular} \caption{Example 3: Here, we consider a multiset of trees $G_m$, which consists of 8 copies of $T_1$, 6 copies of $T_{2}$, 1 copy of $T_{3}$, 7 copies of $T_4$, 7 copies of $T_6$ and 4 copies of $T_{15}$. The total number of 33 is minimal, i.e. $G_m$ is a minimal multiset, such that $T_{1}$ is the unique MDC tree but has no MDC subtrees of size 3 and such that the number of $T_{1}$ in $G_m$ is strictly larger than that of any other $T_i$. Interestingly, {\em all} subtrees of $T_6$ are strictly better than the subtrees of $T_{1}$, but $T_{1}$'s MDC score of 50 is still slightly lower than that of $T_6$, which is 55.}
\label{tab3}
\end{table}

\begin{table}
\begin{tabular}{ |c||c|c|c|c|c|c|  }  \hline $T_i$& $x_i$ & {\scriptsize $l(G_m, T_i)$} & {\scriptsize $l(G_m|_{\{1,2,3\}},T_i|_{\{1,2,3\}})$}&{\scriptsize $l(G_m|_{\{1,2,4\}},T_i|_{\{1,2,4\}})$}&{\scriptsize$l(G_m|_{\{1,3,4\}},T_i|_{\{1,3,4\}})$}&{\scriptsize$l(G_m|_{\{2,3,4\}},T_i|_{\{2,3,4\}})$} \\\hline\hline $T_1$ & 3 & 13 &5&5&{\bf 4}&{\bf 4} \\ \hline $T_2$ & 0 & 15 &5&5&7&6 \\\hline $T_3$ & 0 & 15 &7&5&{\bf 4}&{\bf 4} \\ \hline $T_4$ & 0 &19 &7&7&{\bf 4}&6 \\ \hline $T_5$ &0 &17 &5&7&7&6 \\ \hline $T_6$ & 0 & 19 &7&7&7&6 \\ \hline $T_7$ & 0 & 15 &{\bf 4}&5&{\bf 4}&{\bf 4} \\ \hline $T_8$ & 0 & 15 &{\bf 4}&{\bf 4}&5&{\bf 4} \\ \hline $T_9$ & 0 & 16 &5&{\bf 4}&7&6 \\ \hline $T_{10}$ & 1 & 14 &{\bf 4}&{\bf 4}&5&6 \\ \hline $T_{11}$ & 0 & 18 &7&7&5&6\\ \hline $T_{12}$ & 2 & 14 &{\bf 4}&{\bf 4}&5&6 \\ \hline $T_{13}$ & 0 & {\bf 11} &5&5&5&6\\ \hline $T_{14}$ & 1 & 13 &7&{\bf 4}&{\bf 4}&6 \\ \hline  $T_{15}$ & 1 & 14 &{\bf 4}&7&7&{\bf 4} \\       \hline \end{tabular}\caption{Example 4: Here, we consider a multiset of trees $G_m$, which consists of 3 copies of $T_1$, 1 copy of $T_{10}$, 2 copies of $T_{12}$, 1 copy of $T_{14}$ and 1 copy of $T_{15}$. The total number of 8 is minimal, i.e. $G_m$ is a minimal multiset, such that $T_{13}$ is the unique MDC tree but has no MDC subtrees of size 3 and such that the number of $T_{13}$ in $G_m$ is strictly larger than that of any other $T_i$.  }
\label{tab4} \end{table}



\section{Discussion} We have shown that MDC is not hereditary in the sense that MDC trees need not have any MDC subtrees or MDC triples (note that in our examples with $X=\{1,2,3,4\}$, all subtrees of size $|X|-1$ are triples, so our examples show both non-heredity for subtrees as well as for triples). We showed that this is true for different tree shapes and regardless of the fact if the MDC tree is contained in the input tree set or not. Our proof used the technique of regarding MDC trees simply as the solution of a system of inequalities, which can be easily calculated using the formula on counting the extra lineages needed by a tree to be placed into a species tree presented by \cite{nakleh}. 

However, it is desirable that a method estimating the species tree of a given set of gene trees be hereditary, as the question whether or not a certain taxon is contained in the analysis should not alter the relationship of the remaining taxa. In this sense, non-heredity can be regarded as a drawback of a method.

Our result shows that MDC, which is sometimes regarded as the parsimony method of gene tree reconciliation (\cite{maddison}), suffers from this drawback as does Maximum Parsimony in phylogenetic tree reconstruction (\cite{fischer}). Also, the technique to prove the non-heredity of MDC could also be adapted from an analogous technique for Maximum Parsimony, namely solving a system of inequalities.

On the other hand, non-heredity can also be helpful: \cite{rosenberg} showed that the most likely gene tree on a taxon set $X$ always coincides with the species tree for $|X|=3$, but that this does not in general hold if $|X|\geq 4$. This implies that taking the tree suggested by the majority of gene trees as an estimate for the species tree works for triples, but not necessarily for larger trees. Therefore, the authors suggest to use gene triples and to construct the species tree estimate from these triples by combining them into one common supertree (however, it is well known that there are then other problems like incompatibility of input trees or non-uniqueness of the supertree (c.f. \cite{steel})). So this idea basically says that while estimating the entire species tree by using the majority tree can go wrong, estimating the triples by majority is always correct. In this sense, majority estimates are non-hereditary, too, because the (correct) subtree solution differs from the (possibly incorrect) solution on the entire taxon set -- and \cite{rosenberg} suggest to use this knowledge of non-heredity as a means to overcome the problem of a wrong majority tree estimate by considering triple subtrees instead. 

For MDC, however, it is not so clear if non-heredity has any advantages which could lead to an improved tree reconciliation method, as it is (unlike in the majority scenario) not a priori clear if the subtree estimate or the entire tree estimate is better. This gives rise to a variety of research questions which will be considered in forthcoming papers.

\bibliographystyle{sysbio}
\bibliography{fischer_MDC_bibfile}

\end{document}